\documentclass[conference]{IEEEtran}
\IEEEoverridecommandlockouts
\usepackage{cite}
\usepackage{amsmath,amssymb,amsfonts}
\usepackage{algorithmic}
\usepackage{graphicx}
\usepackage{textcomp}
\usepackage{xcolor}

\usepackage{orcidlink} % Ben added this for the ORCID link
\usepackage{cite} % so we can do \cite{paper1,paper2}

\usepackage{float}     % for the [H] option

\usepackage{braket}
\usepackage{qcircuit}

\newcommand{\figref}[1]{Figure~\ref{#1}}

\def\BibTeX{{\rm B\kern-.05em{\sc i\kern-.025em b}\kern-.08em
    T\kern-.1667em\lower.7ex\hbox{E}\kern-.125emX}}
\begin{document}

\title{Short-time quantum Fourier transform processing\\

\thanks{The research in this paper was carried out during a PhD internship at Dolby Australia. The first author was supported by Sydney Quantum Academy (SQA).
}%as well as the research internship program, Australian Postgraduate Research Intern (APR.Intern).}
}

\author{\IEEEauthorblockN{Sreeraj Rajindran Nair \orcidlink{0009-0003-8753-5621}}
\IEEEauthorblockA{\textit{Dolby Laboratories,} \\
\textit{University of Technology Sydney}\\
Sydney, Australia \\
sreeraj.r.nair@student.uts.edu.au}
\and
\IEEEauthorblockN{Benjamin Southwell \orcidlink{0000-0002-3189-9114}}
\IEEEauthorblockA{%\textit{Advanced Technology Group} \\
\textit{Dolby Laboratories}\\
Sydney, Australia  \\
benjamin.southwell@dolby.com}
\and
\IEEEauthorblockN{Christopher Ferrie \orcidlink{0000-0003-2736-9943}}
\IEEEauthorblockA{%\textit{School of Computer Science} \\
\textit{University of Technology Sydney}\\
Sydney, Australia  \\
christopher.ferrie@uts.edu.au}
}

\maketitle

\begin{abstract}
Algorithms for processing data in short-time batches are critical for both online and offline processing of streamed and large data respectively due to the quadratic relation between signal length and computational cost of convolution-based processing schemes.
%Algorithms for processing data in short-time batches/windows is critical for both online processing in real-time and offline processing of large data, as full-signal processing is often infeasible due to time or resource constraints. 
%since time or resource limitations often make full-signal processing impractical.
%Convolution is a ubiquitous operation used in both classical and quantum algorithms. 
%In sharp contrast to traditional digital signal processing, state-of-the-art in digital quantum signal processing seldom employ the convolution theorem to perform efficient convolutions and instead opt for time-domain convolutions.  
Whilst quantum analogs to some digital signal processing algorithms have been discovered, including the quantum Fourier transform (QFT), there has been no development of short-time processing techniques in the quantum domain.
In this manuscript, we introduce the short-time QFT (STQFT) processing technique to bridge this gap in research. We develop a novel overlap-add reconstruction technique in the quantum domain using a permutation gate to combine subsequent windows. 
With this in mind, we discuss convolution under our novel STQFT processing scheme. 
We demonstrate filtering in the quantum Fourier domain with a filter stored in a quantum register as well as in a block encoded unitary gate. 
%With this in mind, we discuss an application of our novel STQFT processing scheme for filtering in the quantum Fourier domain using the convolution theorem with a filter stored in a quantum register as well as in a block encoded unitary gate. 
Throughout the paper, we elaborate upon implementation details such as applying DC offsets to input signals, skipping input data frames whenever necessary, the use of overlap-save as a reconstruction technique and mitigating time-varying scaling due to normalization of the windowed input data and filters.

\end{abstract}

%%%%%%%%%%%%%%%%%%%%%%%%%%%%%%%%%%%%%%%%%%%%%%%%%%%%%%%%%%%%%%
\section{Introduction}
\label{sec:intro}
% Ben had a very quick rough pass at the introduction. Needs a  lot more but I think it's got the rough bones of what we want.
% Feel free to modify whatever you want however you want
Digital signal processing (DSP) using the short-time Fourier transform (STFT) \cite{allen1977short} is vital for real-time applications and for offline processing of large data in smaller, more manageable chunks, as it is often infeasible to process the
entire signal at once due to time or resource constraints \cite{allenrabiner1977}\cite{DUHAMEL1990259}. Short-time processing is well established within the classical domain, where it transforms  $O(n^2)$
operations into  $O(nlog(n))$ for convolution-based algorithms \cite{stockham}.
The STFT involves windowing a subset of data to be processed, transforming the windowed data into the Fourier domain, processing the signal, applying an inverse Fourier transform to the processed data and then reconstructing the output signal if synthesis is required.
Signal reconstruction is typically done using overlap-add (OLA) or overlap-save \cite{oppenheim1999discrete}.

The STFT is the basis for more advanced analysis and processing techniques that extend the STFT.
This includes the use of a modified discrete Fourier transform (DFT) \cite{karp1999modified} as well as banding and transforming the values of the STFT output, e.g. mel-frequency cepstrum coefficients \cite{sharma2020trends}.
Moreover, processing signals in the sub-bands of the STFT can often lead to more performant algorithms \cite{lee2009subband} in addition to an overall reduction in compute complexity if processing involves operations that scale quickly as the length increases such as convolution.
These benefits are crucial to many applications including machine listening \cite{nassif2019speech, sharma2020trends}, audio \& video coding \cite{vetterli1995wavelets, andersen2004introduction, spanias2006audio} and echo management \cite{lee2009subband, gilloire1992adaptive} for example.

%The traditional classical algorithm for performing short-time Fourier transform convolution involves processing a sliding window of the input signal, performing the convolution by multiplication in the frequency domain and finally reconstructing the signal. There are two approaches for this.  % needs DSP citations for overlap add and overlap save

Recent advances in quantum signal processing (QSP) have resulted in quantum counterparts to traditional DSP algorithms. % unsure if we should throw a citation in here
Of particular significance is the quantum Fourier transform (QFT) \cite{q0, q01, q02} which is the counterpart to the discrete Fourier transform (DFT).
Thus, it is of interest to investigate the short-time QFT (STQFT) to determine if such a processing paradigm can exist in the quantum domain and what the practical implementation issues are.
In this manuscript, we introduce signal processing using the STQFT, present multiple methods for implementing filters in the quantum Fourier domain (QFD), develop a quantum overlap-add (QOLA) algorithm based on a permutation matrix and discuss practical implementation details.
We also note how the overlap-save method is trivially applied to STQFT processing.

\section{Windowing signals in the quantum domain}
\label{sec:quantum_windowing}
%\subsection{teest}
%\subsection{Windowing}

The classical algorithm for performing signal reconstruction involves batching of the input signal into smaller manageable window frames. There are a set of standard windowing functions each with unique characteristics suitable for particular instances \cite{prabhu}. A common window function is the box-car/rectangular window. It generates a subset of the long input signal of the corresponding window length. For the quantum reconstruction algorithm we perform windowing in similar manner. There have been few studies into windowing in the quantum domain \cite{q1, q2, q3}. Quantum windowing delineates from classical windowing as quantum encoding schemes require Euclidean normalization of the windowed signals \cite{nielsen00}. After normalization we encode the windowed signal onto the qubits.

%\subsection{Amplitude Encoding}
Quantum computing offers various forms of data encoding schemes, such as basis encoding and phase encoding  \cite{nielsen00} \cite{dataencod1}. 
%In this paper, we employ amplitude encoding \cite{nielsen00}. This scheme dictates that the individual discrete sample value be embedded directly onto the namesake complex amplitudes of the qubit vector data types. This is a very efficient scheme, as N qubits can accommodate $2^N$ data samples. Reference \cite{q1} proposed creating windowed quantum states with such encoding. 
In this manuscript, we employ amplitude encoding \cite{nielsen00} where the individual discrete sample values are embedded directly onto the namesake complex amplitudes of the qubit vector data types. This is an efficient scheme, as N qubits can accommodate $2^N$ data samples, and it has been used to construct windowed quantum states \cite{q1}.
We thus, are dealing with $2^N$ dimensional Hilbert space. %TODO
% It was proposed to create windowed quantum states with such encoding \cite{q1}.
While encoding, we propose that the windowed signal is padded out beforehand to accommodate for the expected expansion due to the convolution in the subsequent processing. 
The number of qubits is the closest power of 2 over the quantity $w_l + f_l -1$ where $w_l$ is the window length and $f_l$ is the filter length.
%Hence, the number of qubits is determined by the length of window (n) and length of filter (m) put together. In practice this amounts to a vector with length of closest power of 2 over the quantity $n + m -1$.
The quantum state initialised with this is of the form  $\ket{\psi_{window}}=\sum_{i \in {(0,1)^N}}a_i\ket{i}$, where each $a_i$ is a normalized windowed signal sample scalar. In practice, the measurement of this state will reveal the probabilities of each amplitude i.e $a_i^2$, and hence for processing negative amplitudes, a DC offset is applied to the input signal before encoding data. Another resolution would be to modify our algorithm to work with other encoding schemes that build upon amplitude encoding such as Flexible Representation of Quantum Audio (FRQA) \cite{q31}. Furthermore, an all-zero windowed signal cannot be encoded with this scheme and hence its essential to track such states without quantum processing them and instead account for them in the final reconstructed signal as part of classical post-processing.

\section{Filtering in the short-time quantum Fourier domain}
\label{sec:stqft_filtering}
\subsection{Filter as a quantum register}
\begin{figure}[H]
    \centering
    % Adjust the scaling factor (0.8 in this example) as needed
    \scalebox{0.925}{%
        \Qcircuit @C=0.6em @R=0.5em { 
            & \lstick{\ket{0}} & \multigate{2}{\mathbf{Window}} & \multigate{2}{\mathbf{QFT}} & \ctrl{3} & \qw & \qw & \qw & \qw & \multigate{2}{\mathbf{IQFT}}& \meter\\
            %& \ghost{\mathbf{Window}} & \ghost{\mathbf{QFT}} & \qw & \ctrl{4} &\qw & \qw & \qw & \ghost{\mathbf{IQFT}} & \meter\\ 
            & \lstick{\ket{0}} & \ghost{\mathbf{Window}} & \ghost{\mathbf{QFT}}  & & &  &    \vdots   & \nghost{\mathbf{IQFT}} & & \vdots \\ 
            & \lstick{\ket{0}} & \ghost{\mathbf{Window}} & \ghost{\mathbf{QFT}} & \qw & \qw & \qw & \qw & \ctrl{3} &  \ghost{\mathbf{IQFT}} & \meter\\
            & \lstick{\ket{0}} & \multigate{2}{\mathbf{Filter}} & \multigate{2}{\mathbf{QFT}} & \targ & \qw &\qw & \qw & \qw \qw & \meter \\ 
            %& \ghost{\mathbf{Filter}} & \ghost{\mathbf{QFT}} & \qw & \targ & \qw & \qw & \qw \qw & \meter\\ 
            & \lstick{\ket{0}} & \ghost{\mathbf{Filter}} & \ghost{\mathbf{QFT}}  & & &  &    \vdots &   &    \vdots   \\ 
            & \lstick{\ket{0}} & \ghost{\mathbf{Filter}} & \ghost{\mathbf{QFT}} & \qw & \qw & \qw & \qw & \targ &   \meter\\
            }
    }
    \caption{Quantum convolution circuit with filter as a register circuit}
    \label{fig:faaqr}
\end{figure}
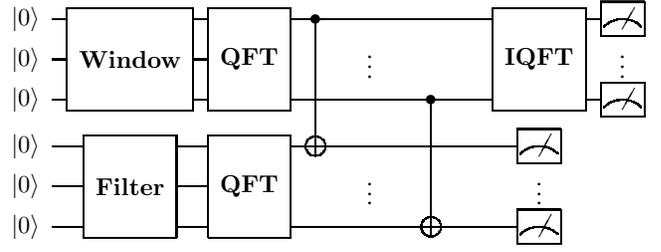

Our first approach for convolution in the quantum domain is with the filter vector of arbitrary length being encoded within a separate quantum register in similar fashion to the amplitude encoding in the case of the windowed signal. 
The filter is padded to the same length that the windowed input data was.
%Care is taken to keep the lengths of vectors and qubit count identical for both the window register as well the filter register. This entails performing identical padding as earlier. 
Once both registers are properly initialized we perform a quantum Fourier transform on both registers separately. With this operation the data is transformed to the Fourier domain which is a crucial part for the application of the convolution theorem \cite{oppenheim1999discrete}. 
%Convolution theorem dictates that convolution in time domain or any simple sample domain is equivalent to the product operation in the Fourier domain. %TODO
To perform elementwise multiplication required for convolution in the quantum Fourier domain we apply repeated controlled pauli X-gate (C-NOT) operations between respective qubits of the two quantum registers \cite{q4}\cite{q5}. 
%Another approach used QFT in tandem with traditional DFT to perform circular convolution \cite{q5}. %TODO
The two registers have the quantum states $\ket{\psi_{window}}=\sum_{i \in {(0,1)^N}}a_i\ket{i}$ and  $\ket{\psi_{filter}}=\sum_{i \in {(0,1)^N}}b_i\ket{i}$. After the application of successive controlled Pauli X-gates (C-NOT) on respective qubits we would have $\ket{\psi_{product}}=\frac{1}{\sqrt{\sum\left|a_ib_i \right|^2}}\sum_{i \in {(0,1)^N}}a_i b_i\ket{i}$. This is inherently a probabilistic operation with the required elementwise product being observed on the first register when the second register is postselected for  $\ket{00...0}$ state. With this postselection put in place the required state will be observed with a probability of $P = \sum\left|a_ib_i \right|^2$.  Subsequently, the first register is subjected to an inverse quantum Fourier transform to obtain the quantum convolution output state.  We used the IBM open-source Python framework Qiskit SDK \cite{q61} to implement this scheme and found that it matches the classical DSP convolution output within the tolerance of machine precision. %TODO
The general circuit construction of this scheme is shown in \figref{fig:faaqr}.

% \begin{figure}[H] % Use [H] command to force latex to place figures within text where its placed in the code rather than float around
% % INSEERT qccircuiitcode here
% \Qcircuit @C=0.6em @R=0.5em { 
% & \lstick{\ket{0}} & \multigate{2}{\mathbf{Window}} & \multigate{2}{\mathbf{QFT}} & \ctrl{3} & \qw & \qw & \qw & \qw & \multigate{2}{\mathbf{IQFT}}& \meter\\
% %& \ghost{\mathbf{Window}} & \ghost{\mathbf{QFT}} & \qw & \ctrl{4} &\qw & \qw & \qw & \ghost{\mathbf{IQFT}} & \meter\\ 
% & \lstick{\ket{0}} & \ghost{\mathbf{Window}} & \ghost{\mathbf{QFT}}  & & &  &    \vdots   & \nghost{\mathbf{IQFT}} & & \vdots \\ 
% & \lstick{\ket{0}} & \ghost{\mathbf{Window}} & \ghost{\mathbf{QFT}} & \qw & \qw & \qw & \qw & \ctrl{3} &  \ghost{\mathbf{IQFT}} & \meter\\
% & \lstick{\ket{0}} & \multigate{2}{\mathbf{Filter}} & \multigate{2}{\mathbf{QFT}} & \targ & \qw &\qw & \qw & \qw \qw & \meter \\ 
% %& \ghost{\mathbf{Filter}} & \ghost{\mathbf{QFT}} & \qw & \targ & \qw & \qw & \qw \qw & \meter\\ 
% & \lstick{\ket{0}} & \ghost{\mathbf{Filter}} & \ghost{\mathbf{QFT}}  & & &  &    \vdots &   &    \vdots   \\ 
% & \lstick{\ket{0}} & \ghost{\mathbf{Filter}} & \ghost{\mathbf{QFT}} & \qw & \qw & \qw & \qw & \targ &   \meter\\
% }
% \caption{Quantum Convolution circuit with Filter as a Register Circuit}
% \label{fig:faaqr}
%\end{figure}

%%%%%%%%%%%%%%%%%%%%%%%%%%%

%%%%%%%%%%%%%%%%%%%%%%%%%%

This convolved quantum state can be matched to the classical convolution output by rescalling the classical read out of the quantum circuit with the normalization factors and the probability factors accrued during the operations. Filter as a quantum register approach would require $O(2^N)$ iterations to produce reliable results with further iterations to map out all the amplitudes of the quantum state. Existing work relates quantum amplitude amplification with the filter as a register scheme to produce more efficient run counts \cite{q4}. %%%

\subsection{Filter as a block encoded matrix}
    A novel approach to perform same aforementioned elementwise multiplication is to use unitary operations in place of a separate quantum register. To carry out this operation, we need to be able to embed the filter vector as a diagonal matrix. Any other configurations are likely prone to introducing errors into the convolution output. The challenge with this approach is that such a diagonal matrix would not be of unitary nature. A resolution for this is to instead embed the filter vector as the diagonal block matrix within a much larger unitary matrix. 
    %with ancilla qubits to handle the expanded Hilbert space dimensions. %TODO
    %Efficient quantum gate decomposition of such a larger unitary matrix is not always guaranteed. 

Block encoding \cite{q7} is one such scheme for embedding a properly scaled non-unitary matrix $A_{i,j}\in C$ onto a unitary matrix $U_{Block}$ :

\begin{equation}
    U_{Block} = \begin{bmatrix}
            A_{i,j} & * \\
            * & *
            \end{bmatrix}
\end{equation}
Here $*$ denotes the arbitrary blocks we have no interest in. The action of this block encoded matrix on a general quantum state with ancilla qubits of the form $\ket{\psi}= \ket{0}\ket{\alpha} $ is as follows:

\begin{align}
\label{eqn:eqlabel}
\begin{split}
     U_{Block}\ket{\psi} &= \begin{bmatrix}
            A_{i,j} & * \\
            * & *
            \end{bmatrix}
            \begin{bmatrix}
            \alpha \\
            \mathbf{0}
            \end{bmatrix}=
            \begin{bmatrix}
            A_{i,j}\alpha \\
            *
            \end{bmatrix} 
\\
 &= \ket{0}(A_{i,j}\ket{\alpha}) + \ket{1}\ket{*}
\end{split}
\end{align}

\begin{figure}[H]
    \centering
    \scalebox{0.95}{%
        \Qcircuit @C=1.9em @R=1.2em { 
            & \lstick{\ket{0_{anc}}} & \qw & \qw & \multigate{2}{\mathbf{U_A}} & \qw & \qw & \qw \\ 
            & \lstick{\ket{i}} & {/} &  \gate{H^{\otimes n}}  & \ghost{\mathbf{U_A}} & \multigate{1}{\mathbf{U_B}} & \gate{H^{\otimes n}} & \qw \\
            & \lstick{\ket{j}} & {/} & \qw & \ghost{\mathbf{U_A}} & \ghost{\mathbf{U_B}} & \qw & \qw 
        }
    }
    \caption{Oracle access block encoding circuit}
    \label{fig:bec}
\end{figure}
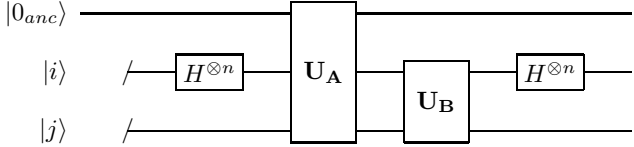

We propose the use of a matrix access oracle based block encoding scheme \cite{q7}\cite{q8}. The general structure of such a scheme as a quantum circuit is shown in \figref{fig:bec}. $\mathbf{U_A}$ and $\mathbf{U_B}$ are oracles responsible for encoding the matrix elements of our diagonal matrix $A_{i,j}$ and ensure iteration over all combinations of the indices $i,j$ respectively. %(for our purposes we only need cases where $i=j$).
A bottleneck in this scheme is that an efficient quantum gate decomposition for $\mathbf{U_A}$ and $\mathbf{U_B}$ may not always be found. 

The oracles' $\mathbf{U_A}$ and $\mathbf{U_B}$ operation on a quantum state is defined as follows:

\begin{align}
\label{eqn:eqlabel}
\begin{split}
 U_{A}\ket{0}_{anc}\ket{i}\ket{j} &= \ket{A_{i,j}}_{anc}\ket{i}\ket{j} ,
\\
 U_{B}\ket{i}\ket{j} &= \ket{j}\ket{i} .
\end{split}
\end{align}

where $\ket{A_{i,j}}_{anc} \equiv A_{i,j}\ket{0}_{anc}+\sqrt{1-\left|A_{i,j} \right|^{2}}\ket{1}_{anc}$. %incorrect "=" instead of "+") and is rectified this in the revision.

We use a more efficient scheme built on top of the oracle access approach known as the Fast Approximate BLock Encodings (FABLE) \cite{q7}. FABLE uses a combination of single  $R_y$  rotation ad controlled pauli X-gates (C-NOT) to construct the circuits for the oracles.
%For our experiments we used the open-source Python framework Pennylane SDK \cite{q9} to generate these FABLE matrices and gate decompositions. 
The general circuit for these constructions is shown in \figref{fig:bec}. 
%These gate decompositions were then used within a local quantum simulator to perform windowed quantum convolutions and found it matches the traditional DSP convolution output. %TODO

% \begin{figure}[H] % Use [H] command to force latex to place figures within text where its placed in the code rather than float around
% % INSEERT qccircuiitcode here
% \centering
% \Qcircuit @C=1.9em @R=1.2em { 
% & \lstick{\ket{0_{anc}}} & \qw & \qw & \multigate{2}{\mathbf{U_A}} & \qw & \qw & \qw \\ 
% & \lstick{\ket{i}} & {/} &  \gate{H^{\otimes n }}  & \ghost{\mathbf{U_A}} & \multigate{1}{\mathbf{U_B}} \qw & \gate{H^{\otimes n }} & \qw \\
% & \lstick{\ket{j}}  & {/} & \qw & \ghost{\mathbf{U_A}} & \ghost{\mathbf{U_B}} \qw & \qw & \qw 
% } 
% \caption{Oracle Access Block Encoding Circuit}
% \label{fig:bec}
% \end{figure}

%%%%%%%%%%%%%%%%%%%%%%%%%%%%

Before we can encode our diagonal $A$ matrix we must transform the diagonal elements to the frequency domain. This is performed by finding the QFT coefficients of the filter with a separate circuit before passing these to the matrix $A$. As before we pad the length of the diagonal with zeros to account for the anticipated expansion to length $n \geq w_l + f_l -1$ due to the convolution in the subsequent steps. We now have a $n \times n$ filter block within the unitary matrix. This will operate on $log_2 n$ primary qubits and additional ancilla qubits.  %TODO

\[
U_{Block} = \begin{bmatrix}
 \begin{bmatrix}
   a_{1,1} & 0 & 0 & \dots & 0 & 0\\
   0 & \ddots & 0 & \dots & 0 & 0\\
   0 & 0 & a_{f_l,f_l} & \ddots & \vdots & \vdots\\
   \vdots & \vdots & \ddots & 0 & 0 & 0\\
   0 & 0 & \dots & 0 & \ddots & 0\\
   0 & 0 & \dots & 0 & 0 & 0\\
  \end{bmatrix} &   \mathbf{*}          \\
      \mathbf{*}  & \mathbf{*}  &        \\
   %   \textbf{*}  & \mathbf{*}  &        \\      
\end{bmatrix}
\]

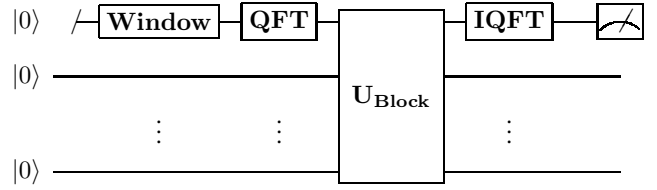
\begin{figure}[H]
    \centering
    % Replace 0.8 below with whatever scaling factor you need
    \scalebox{0.95}{%
        \Qcircuit @C=0.9em @R=1.0em {
            & \lstick{\ket{0}} & {/} & \gate{\mathbf{Window}} & \gate{\mathbf{QFT}} & \multigate{3}{\mathbf{U_{Block}}} & \gate{\mathbf{IQFT}} & \qw & \meter \\
            & \lstick{\ket{0}} & \qw & \qw & \qw & \ghost{\mathbf{U_{Block}}} & \qw & \qw & \qw \\
            & & & \vdots & \vdots & \nghost{\mathbf{U_{Block}}} & \vdots & & \\
            & \lstick{\ket{0}} & \qw & \qw & \qw & \ghost{\mathbf{U_{Block}}} & \qw & \qw & \qw \\
        }
    }
    \caption{Quantum convolution circuit with filter as a block matrix}
    \label{fig:faabm1}
\end{figure}

This block encode gate decomposition is applied to the Fourier transformed windowed signal encoded state. Ancilla qubits are introduced to accommodate the large operational Hilbert space of the $U_{Block}$ unitary operator. Finally, an inverse quantum Fourier transform is performed on this state to obtain the convolution output for the windowed signal.
For our experiments we used the open-source Python framework Pennylane SDK \cite{q9} to generate the FABLE matrices and their gate decompositions.
These gate decompositions were then used within a local quantum
simulator to perform windowed quantum convolutions and
found it matches the traditional DSP convolution output within the tolerance of machine precision.
The general structure of the circuit is shown in \figref{fig:faabm1}.

\section{Quantum overlap-add}
\label{sec:qola}
Classical reconstruction of a processed windowed signal is widely performed using either overlap-add or overlap-save. To the best of the authors' knowledge, a quantum version of these sub-routines has not been developed. In this section we propose a reconstruction algorithm in the form of a quantum overlap-add (QOLA) circuit. 

For QOLA, we collect the classical readouts of windowed quantum convolution outputs. In general, these would be of identical lengths. After rescaling to account for the normalization factors and probability terms, these are zero padded to the closest $2^N$ length if required. A pair of consecutive window convolution outputs $\ket{\phi_a}$ and $\ket{\phi_b}$ each of length $r$ are amplitude encoded together into a $log_2(2r)$ qubit register to create the state $\ket{\phi_{QOLA}} = \ket{0}\ket{\phi_{a}} + \ket{1}\ket{\phi_{b}}$ with an additional euclidean norm.
%A $\ket{\phi_{QOLA}}$ can be decomposed as $\ket{\phi_{QOLA}} = \ket{0}\ket{\phi_{a}} + \ket{1}\ket{\phi_{b}}$.

%\[
%\ket{\phi_{QOLA}} = \ket{0}\ket{\phi_{a}} + \ket{1}\ket{\phi_{b}}
%\]
As was the case with the windowed quantum state initialization, an all-zero pair of windowed convolution outputs cannot be processed with this scheme and must be relegated to the classical post-processing wherein we insert all-zero frames between reconstructed output frames.
%The two consecutive windowed convolution outputs are split into two terms as $\ket{\phi_a}$ and $\ket{\phi_b}$.
% A controlled NOT operation is performed between an ancilla qubit and the $1^{st}$ qubit in $\ket{\phi_{QOLA}}$ to give new decomposition terms $\ket{\phi_a^'}$ and $\ket{\phi_b^'}$. This gives the state:
A controlled-NOT operation is applied between an ancilla qubit and the
first qubit in the state $\ket{\phi_{\mathrm{QOLA}}}$, producing new
decomposition terms $\ket{\phi_a'}$ and $\ket{\phi_b'}$.  
The resulting state is

% \begin{align}
% \label{eqn:eqlabel}
% \begin{split}
% \ket{0}\ket{\phi_{QOLA}} = &\ket{0}\ket{0}\ket{\phi_{a}} + \ket{1}\ket{1}\ket{\phi_{b}} \\
% = &\ket{0}\ket{\phi_{a}^'} + \ket{1}\ket{\phi_{b}^'} \\
% = & \ket{0} \otimes [\phi_1 , \phi_2 , ... \phi_r , 0, 0, ... , 0]
% \\
% + & \ket{1} \otimes [0,  0, ... ,     0, \phi_{r + 1} , \phi_{r + 2} , ... \phi_{2r}  ]
% \end{split}
% \end{align}

\begin{align}
\ket{0}\ket{\phi_{\mathrm{QOLA}}}
  &= \ket{0}\ket{0}\ket{\phi_a} \;+\; \ket{1}\ket{1}\ket{\phi_b} \nonumber\\
  &= \ket{0}\ket{\phi_a'}        \;+\; \ket{1}\ket{\phi_b'} \nonumber\\
  &= \ket{0}\!\otimes\!\bigl[\phi_1,\phi_2,\ldots,\phi_r,0,0,\ldots,0\bigr] \nonumber\\
  &\quad+\;\ket{1}\!\otimes\!\bigl[0,0,\ldots,0,\phi_{r+1},\phi_{r+2},\ldots,\phi_{2r}\bigr].
\label{eq:eqlabel}
\end{align}

% Sreeraj, I don't think it's good to specify np.roll() which is a function of a particular library. You should talk about the generalised operation. This is a circular shift https://en.wikipedia.org/wiki/Circular_shift
% This is a good approach for everything (so it applies to Qiskit / pennylane etc whenever a generalised function is being used.
% Be specific about how the circular shift is applied. (e.g. 
An arbitrary ancilla qubit controlled permutation matrix ($C-U_{PERM}$) reorganizes the coefficients in $\ket{\phi_b}$ to the desired overlap ratio configuration. 
%One way to generate the $U_{PERM}$ matrix is to create an appropriately sized identity matrix and morphing it with circular shift operations. 
We define a $2r \times 2r$ permuation matrix for $l$ overlapping elements:

%\begin{equation}
%    U_{PERM} = \begin{bmatrix}
%            I_{l\times l} & \mathbf{0}_{l\times l} & \mathbf{0}_{l\times z} \\
%            \mathbf{0}_{w \times l } & \mathbf{0}_{w \times l }& I_{z\times z} \\
 %           \mathbf{0}_{l\times l} & I_{l\times l} &  \mathbf{0}_{l \times z } \\
 %           \end{bmatrix}
%\end{equation}

\begin{equation}
    U_{PERM} = \begin{bmatrix}
            I_{r-l\times r-l} & \mathbf{0}_{r-l\times l} & \mathbf{0}_{r-l\times r} \\
            \mathbf{0}_{r \times r-l } & \mathbf{0}_{r \times l }& I_{r\times r} \\
            \mathbf{0}_{l\times r-l} & I_{l\times l} &  \mathbf{0}_{l \times r } \\
            \end{bmatrix}_{2r \times 2r}
\end{equation}
%where $l$ is the number of overlapping elements.

%where $l$ is the overlap size, $w = z - 2l$ and $z = 2^n$ i (CHANGE THIS $n$ SOMEWHERE). 
$U_{PERM}$ is then converted into the controlled permutation gate ($C-U_{PERM}$).

%$C-U_{PERM}$ \in \Re{}

\begin{align}
\label{eqn:eqlabel}
\begin{split}
\xrightarrow{C-U_{PERM}}  & \ket{0} \otimes [\phi_1 , \phi_2 , ... \phi_r , 0, 0, ... , 0]
\\
+ & \ket{1} \otimes [0,  0, ... ,     0, \phi_{r + 1} , \phi_{r + 2} , ... \phi_{2r} , 0, 0, ... , 0 ]
\end{split}
\end{align}

A Hadamard operation on the ancilla qubit initiates the addition operation. A postselection of $\ket{0}$ on the ancilla qubit implies a quantum overlap-add state on the first register. On the other hand $\ket{1}$ on the ancilla qubit implies a difference state on the first register.

% \begin{align}
% \label{eqn:eqlabel}
% \begin{split}
% \xrightarrow{Hadamard}  & \frac{1}{\sqrt{2}} \ket{0} \otimes [\ket{\phi_a ^'} + \ket{\phi_b^'}]
% \\
% + & \frac{1}{\sqrt{2}} \ket{1} \otimes [\ket{\phi_a^'} - \ket{\phi_b^'}]
% \\
% = & \frac{1}{\sqrt{2}} \ket{0} \otimes \ket{SUM}
% \\
% + & \frac{1}{\sqrt{2}} \ket{1} \otimes \ket{DIFFERENCE}
% \end{split}
% \end{align}

\begin{align}
\xrightarrow{\text{Hadamard}}\;
 &\frac{1}{\sqrt{2}}\,
   \ket{0}\!\otimes\!\bigl(\ket{\phi_a'}+\ket{\phi_b'}\bigr)
 +\frac{1}{\sqrt{2}}\,
   \ket{1}\!\otimes\!\bigl(\ket{\phi_a'}-\ket{\phi_b'}\bigr) \nonumber\\
 ={}&\frac{1}{\sqrt{2}}\,
   \ket{0}\!\otimes\!\ket{\mathrm{SUM}}
 +\frac{1}{\sqrt{2}}\,
   \ket{1}\!\otimes\!\ket{\mathrm{DIFFERENCE}}
\label{eq:eqlabel}
\end{align}

% Both these measurements occur with probabilities of $\frac{1}{2}\left|\ket{SUM} \right|^2$ and $\frac{1}{2}\left|\ket{DIFFERENCE}\right|^2$ respectively . The classical readouts are rescaled with these probability and the normalization factors and then we obtain the required output. 
Both outcomes occur with probabilities
\(\tfrac12\lVert\mathrm{SUM}\rVert^{2}\) and
\(\tfrac12\lVert\mathrm{DIFFERENCE}\rVert^{2}\), respectively.
The classical read-outs are then rescaled by these probabilities (and any
overall normalisation factor) to produce the desired output.

\begin{figure}[H]
    \centering
    \scalebox{0.9}{%
        \Qcircuit @C=1.9em @R=0.8em {
            & \lstick{\ket{0}} & \multigate{2}{\mathbf{Encode}} & \ctrl{3} & \multigate{2}{\mathbf{U_{PERM}}} & \qw & \qw \\
            & \lstick{\ket{0}} & \ghost{\mathbf{Encode}}       & \qw      & \ghost{\mathbf{U_{PERM}}}         & \qw  & \qw\\
            & \lstick{\ket{0}} & \ghost{\mathbf{Encode}}       & \qw      & \ghost{\mathbf{U_{PERM}}}         & \qw & \qw\\
            & \lstick{\ket{0_{anc}}} & \qw   & \targ    & \ctrl{-1} & \gate{H} & \meter
        }
    }
    \caption{Quantum overlap-add circuit for (3+1) qubits}
    \label{fig:qola}
\end{figure}
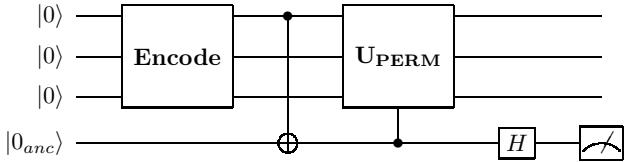

In general, each windowed convolution output undergoes overlap-add twice, one with the preceding windowed convolution output and one with the succeeding one. We replicated these results with Qiskit and found it matches the traditional OLA within the tolerance of machine precision. The (3+1) qubit circuit for this operation is shown in \figref{fig:qola}.

% \begin{figure}[H] % Use [H] command to force latex to place figures within text where its placed in the code rather than float around
% \Qcircuit @C=1.9em @R=0.8em {
% & \lstick{\ket{0}} & \multigate{2}{\mathbf{Encode}} & \ctrl{3} &  \multigate{2}{\mathbf{U_{PERM}}} & \qw \\ 
% & \lstick{\ket{0}} & \ghost{\mathbf{Encode}} & \qw &  \ghost{\mathbf{U_{PERM}}} & \qw\\ 
% & \lstick{\ket{0}} & \ghost{\mathbf{Encode}} & \qw  & \ghost{\mathbf{U_{PERM}}} & \qw \\ 
% & \lstick{\ket{0_{anc}}} & \qw & \targ  & \ctrl{-1} &   \gate{H} & \meter \\
% }
% \caption{Quantum Overlap Add Circuit for (3+1)-qubits}
% \label{fig:qola}
% \end{figure}

%%%%%%%%%%%%%%%%%%%%%%%%%%%%%%%

%%%%%%%%%%%%%%%%%%%%%%%%%%%%5%

\section{Discussion}

%COMMENT: fill this space with bottlencks limitations and future scope for expansion of the work. maybe have conclusion here or do i need separate section. Got rid of experiments section, we talk abt simulations within the main body itself, doesn't really warrant a separate section. Choose between the two filter as a block figures.

% BLOCK ENCODING 
The oracle access based block encoding scheme was found more adept at generating efficient block encodings for the particular filter matrices we are interested in. Investigations into the linear combination of unitaries (LCU) based block encoding schemes didn't yield favorable results for our requirements. It is worth noting that the standard oracle access scheme we demonstrated does not scale efficiently in terms of processing time or circuit depth w.r.t the filter length. We observed that it was better to generate these block encodings for standard filters beforehand. We then call these encodings in, when we perform real-time processing. FABLE does offer more optimal block encodings in the case of certain classes of sparse and/or structured block matrices. However, our required filter matrices did not fit into these classes. In the same vein, it is worth investigating if the structure of the required filter matrices can be exploited to generate efficient block encodings. 
Furthermore, instead of processing the filter array with a QFT circuit separately, we can directly block encode the filter as such and use techniques from QSP \cite{q101} and quantum singular value transform (QSVT) \cite{q102} to transform the filter block to the Fourier domain. This is likely to incur penalties in the form of additional ancilla qubits.
%We process the filter array with a QFT circuit before block encoding it. It is possible to skip this step and directly block encode the filter as such and use techniques from QSP \cite{q101} and Quantum singular value transform (QSVT) \cite{q102} to transform the filter block to the Fourier domain. This is likely to incur penalties in the form of additional ancilla qubits.

% QOLA
The QOLA method accepts inputs as classical data and performs re-encoding of the windowed convolution outputs. It is possible to design QOLA circuits to accept windowed quantum convolution states as input and generate reconstructed frames. However, given that each of the windowed quantum convolution outputs have different norms means that the reconstructed frames would come out scaled unevenly. One workaround is to introduce an arbitrary scalar factor multiplier quantum circuit as a buffer between these two sub-routines, however it is not clear if this can be done without introducing new norm scales along the way. If this can be achieved, either a pair of convolution circuits can be used simultaneously or a single convolution circuit can be used to generate at least two set of output states for each window frame. In later case these states can then be temporarily stored in quantum memory elements such as QRAM \cite{q103} before passing it to the QOLA circuit. This is likely to cause latency issues when deployed to a real-time processing environment. It is tantamount that this sub-routine is optimized to handle these anticipated latency issues.

% OVERLAP_SAVE
Overlap-save is yet another popular signal reconstruction scheme widely used in DSP. We do not need to design a separate quantum circuit in order to adapt this scheme to the quantum data domain. It uses windowed circular convolution outputs instead of linear convolution outputs and a snipping of the output data. A simple end-to-end stacking of these frames constitutes an overlap-save reconstructed signal. This entails that there is no zero padding to $2^N$ over $w_l+f_l-1$ in the case of quantum convolution. We replace the non-overlapping windows from our scheme with overlapping windows, where filter length determines the overlap ratio. The convolved output wraps around in the final windowed quantum convolution state. Classical read-outs are obtained and these are then processed exactly as in traditional DSP.

\section{Conclusion}
In this manuscript, we introduced a novel algorithm for processing signals within the short-time quantum Fourier domain. 
%We develop filtering under our short time quantum Fourier transform processing. 
Quantum filtering sub-routines were designed with two different approaches: one where the filter was encoded as a quantum register and the other where it was encoded as a block encoded unitary gate. 
The convolved windowed signal output frames obtained from these quantum convolutions were then reconstructed with the novel quantum overlap-add technique we proposed. We discussed the current bottlenecks in our approach with the critical one being the classical intermediates between the windowed convolution sub-routine and the quantum overlap-add subroutine. These and several other avenues for improvement to our current scheme were discussed. We foresee STQFT processing to have applications in real-time signal processing such as live streaming and in efficient processing of large data in the near future.

%The quantum convolved output frames for short-time windowed input signal obtained from these convolutions were then stitched back together with Quantum Overlap-Add to generate the final reconstructed signal.

%In this manuscript, we introduced a novel algorithm to perform quantum convolution with windowed signals using the convolution theorem. Quantum convolution sub-routines were implemented in two ways: one where the filter is encoded within a quantum register and second where it was encoded within a block encoded gate. The quantum convolved output frames retrieved from these convolutions were then stitched back together using Quantum Overlap-Add to generate the final reconstructed signal. This algorithm is expected to have applications in real-time signal processing such as live streaming in the foreseeable future.

\section*{Acknowledgment}
S.R.N would like to thank Afrad Basheer for valuable discussions. 
%He would also like to acknowledge his doctoral research funding agency, Sydney Quantum Academy (SQA) as well as the research internship program, Australian Postgraduate Research Intern (APR.Intern).

%\newpage

\bibliographystyle{IEEEtran}
\bibliography{IEEEabrv,references}
 
\vspace{12pt}

% Algorithms for processing data in short-time batches are critical for both online and offline processing of streamed and large data respectively due to the quadratic relation between signal length and computational cost of convolution-based processing schemes. Whilst quantum analogs to some digital signal processing algorithms have been discovered, including the quantum Fourier transform (QFT), there has been no development of short-time processing techniques in the quantum domain. In this manuscript, we introduce the short-time QFT (STQFT) processing technique to bridge this gap in research. We develop a novel overlap-add reconstruction technique in the quantum domain using a permutation gate to combine subsequent windows. With this in mind, we discuss convolution under our novel STQFT processing scheme. We demonstrate filtering in the quantum Fourier domain with a filter stored in a quantum register as well as in a block encoded unitary gate. Throughout the paper, we elaborate upon implementation details such as applying DC offsets to input signals, skipping input data frames whenever necessary, the use of overlap-save as a reconstruction technique and mitigating time-varying scaling due to normalization of the windowed input data and filters.

\end{document}